# Trends in bonding configuration at SiC/III-V semiconductor interfaces

Jin-Cheng Zheng[a)], Hui-Qiong Wang[b)], A. T. S. Wee and C. H. A. Huan
Department of Physics, National University of Singapore, Lower Kent Ridge Road,
Singapore 119260

a) Current mailing address: Theory of Condensed Matter, Cavendish Laboratory, Madingley Road, Cambridge CB3 0HE, United Kingdom.
Email: jincheng_zheng@yahoo.com (J-C Zheng)
b) Current address: Applied Physics, Yale University, Becton Center, 15 Prospect Street, New Haven, CT 06520, USA.

**Abstract**
    The structural and electronic properties of interfaces between β-SiC and III-V semiconductors are studied by first-principles calculations. Favorable bonding configurations are found to form between Si-V and C-III (model A) for BN, AlN, AlP, AlAs, GaN, GaP, GaAs, InN, InP, InAs and InSb, and Si-III and C-V (model B) for BP, BAs, BSb, AlSb and GaSb. The relationship between formation energy difference and lattice constant difference as well as charge distribution for these two models is found. The origin of bonding configurations can be explained in terms of the ionicity of III-V semiconductors, electrostatic effect, charge distribution and band-structure component.

    There has been increasing interest in silicon carbide (SiC) due to its favorable electronic properties, anomalous charge transfer, and extreme elastic and thermal properties[1-5]. The technological realization of self-aggregating wires[6,7] and quantized homostructures[8] make it one of the most promising materials for nanodevices, microelectronics, sensors, and high-power, high-temperature devices. An understanding of the SiC/substrate interface is important for the growth of high quality SiC films. Furthermore, SiC is a promising substrate material for the growth of GaN or AlN semiconductors since GaN and AlN are both well lattice-matched to SiC. Previous studies on the interfaces between SiC and nitrides such as BN[9], AlN[10-13], GaN[13-15] as well as BP[10] semiconductors have revealed the favorable bonding configuration to be Si-N and C-B(Al,Ga) rather than Si-B(Al,Ga) and C-N. In contrast, the stable bonding configuration of SiC/BP is Si-B & C-P instead of Si-P & C-B[10]. BP may be an example of "anomalous" bonding configuration, whilst BN, AlN and GaN are typical examples with "normal" bonding configurations. Several questions arise as to why the bonding configuration between SiC and BP is different from that between SiC and nitrides. A systematic study of bonding configurations at SiC/(III-V) semiconductor interfaces is therefore needed to address these questions. In this work, the linear-muffin-tin-orbital (LMTO) band structure method[16] and local-density-functional (LDA) theory are used for electronic structure and total-energy calculations. We first study the lattice constant and total energy of (001) interfaces of SiC/(III-V), and then discuss the general trends in bonding configuration for such interfaces.

In our calculations, LMTO[16] is used in the atomic-sphere approximation (ASA). The approach is based on the Hohenberg-Kohn-Sham density-functional method in the local-density approximation[17]. To ensure an adequate description of the potential at the tetrahedral interstitial sites, so-called empty spheres[18] are introduced at suitable sites while preserving the crystal symmetry. It has been well established that ASA with interstitial empty spheres gives a complete description of the electronic states and ground-state properties in bulk semiconductors and at semiconductor interfaces[19,20]. The supercell approach is employed to calculate the electronic structure and properties of SiC/(III-V) superlattices, and to compare two different bonding configurations, i.e. Si-V and C-III for Model A, and Si-III and C-V for Model B. The SiC/III-V superlattice (1+1) structure consists of periodic alternating layers of SiC and III-V semiconductors repeating in the (001) direction, as shown in Fig 1. This structure is a special case since it is both a (1+1) (001) superlattice and a (110) superlattice. It is also the CuAu structure (L1$_0$ structure) in $(SiC)_x(III-V)_{1-x}$ alloys with composition x=0.5. The phase diagram (stability) of $(SiC)_x(III-V)_{1-x}$ alloys can be calculated from the formation energy of SiC/III-V structure by a cluster expansion[21,22], which is a generalization of the Connolly–Williams approach[23]. Moreover, the superlattice (1+1) is the simplest structure to distinguish the different bonding configuration of models A and B.

The total energy for SiC and 16 III-V semiconductors were calculated and their bulk equilibrium lattice constants obtained. The formation energies (see Ref.9 for definition) of the (1+1) superlattice of SiC/III-V semiconductors were then calculated to compare the different bonding configurations of model A (Si-V, C-III) and model B (Si-III, C-V).

The formation energy difference (i.e., $\Delta E_{form}(B-A) = E_{form}(B) - E_{form}(A)$) of models A and B of SiC/(III-V) superlattice (1+1) along (001) are presented in Fig 2 (a). We find that model A (Si-V, C-III) is stable for SiC/(III-V) with (III-V) = BN; AlN, AlP, AlAs; GaN, GaP, GaAs; InN, InP, InAs, InSb. These results are in agreement with our previous studies on SiC/BN[9], and other works on SiC/AlN[10-13], and SiC/GaN[13-15]. For SiC/AlP, SiC/AlAs; SiC/GaP, SiC/GaAs, SiC/InN, SiC/InP, SiC/InAs and SiC/InSb, we predict that they prefer the Si-V & C-III (model A) bonding configuration. In previous studies, the bonding configuration at the SiC/BP interface was predicted to be Si-B & C-P (model B) due to the "anomalous" ionicity of BP (B is anion, and P is cation)[10]. This study confirms that the favorable bonding configuration of SiC/BP is indeed Si-B & C-P. Interestingly, we find that besides SiC/BP, several other SiC/(III-V) interfaces prefer model B configurations, namely SiC/BAs, SiC/BSb, SiC/AlSb and SiC/GaSb. We note that the SiC/(III-V) formation energy of model A increases, while that of model B decreases, as the group V element changes from "N" to "Sb". The formation energy difference between models A and B decreases as the group V element changes from "N" to "Sb", as shown in Fig 2 (a).

The formation energy difference and lattice constant difference has a linear relationship, as shown in Fig 3: $\Delta E(eV/atom) = 0.2\Delta a^{B-A}$ where $\Delta a^{B-A} = [a_0(B) - a_0(A)]/a_0(I) \times 100$. Here $a_0(B)$ is the lattice constant of SiC/(III-V) superlattice (1+1) with model B, $a_0(A)$ is that with model A, and $a_0(I)$ is that of the ideal case [average of SiC and (III-V) bulk lattice constants]. The lattice constant of the stable compound is close to the ideal lattice constant, while that of the unstable compound expands, i.e., $a(unstable) > a(stable) > a(ideal)$. This can be explained by the large electrostatic energy in the unstable bonding configurations of SiC/(III-V) that generate repulsive forces causing the lattice to expand. The linear relationship between formation energy difference and lattice constant difference indicate that the degree of instability of the bonding configuration is linearly related to the degree of lattice expansion.

The origin of bonding configurations can be explained in terms of the ionicity of III-V semiconductors, electrostatic effects, charge distribution, and band-structure component. Regarding electrostatic effects, the cation-anion bonding is expected to be stable for bonding configuration at SiC/(III-V) interfaces. The charge distributions of III-V semiconductors reflect their ionicity and clearly relate to the bonding configurations, as shown in Fig 2 (b). We note that in SiC, Si acts as the cation (+1.166 |e| positive charge) and C as the anion (-1.166 |e| negative charge). Because the lower valence levels are associated with the carbon atom, it plays the role of anion. From electrostatic arguments, cations prefer to bond to anions at the interface between two compounds. The charge distribution in III-V semiconductors is clearly consistent with the predicted bonding configurations at the SiC/(III-V) interface except for InSb. InSb has low ionicity (the charge distribution in In is about −0.03, and that of Sb is about +0.03 |e|), and hence, it is not surprising that the energy difference between models A (Si-Sb, C-In) and B (Si-In, C-Sb) is small. The total energy calculations give a small difference of 0.0040 eV/atom. Other effects such as strain energy will affect the bonding configuration of SiC/InSb since this interface has the largest lattice mismatch.

The other important effect comes from band-structure components. When the bonding configuration is unfavorable, localized interface states occur in the main band gap, pushing the Fermi level up and causing the bands to shift upward. For example, SiC/BN with favorable bonding configuration (Si-N, C-B) shows semiconductor characteristics, while the unfavorable model (Si-B, C-N) exhibits anomalous metallic properties[9].

In summary, total-energy calculations have been performed for SiC/III-V semiconductors, and the general trends of the equilibrium lattice constant, formation energy and, bonding configuration are obtained. The relationship between the bonding configuration at the SiC/III-V interfaces and bulk charge distribution in SiC and III-V semiconductors, as well as that between formation energy and lattice constant, is discussed.

**List of figures:**

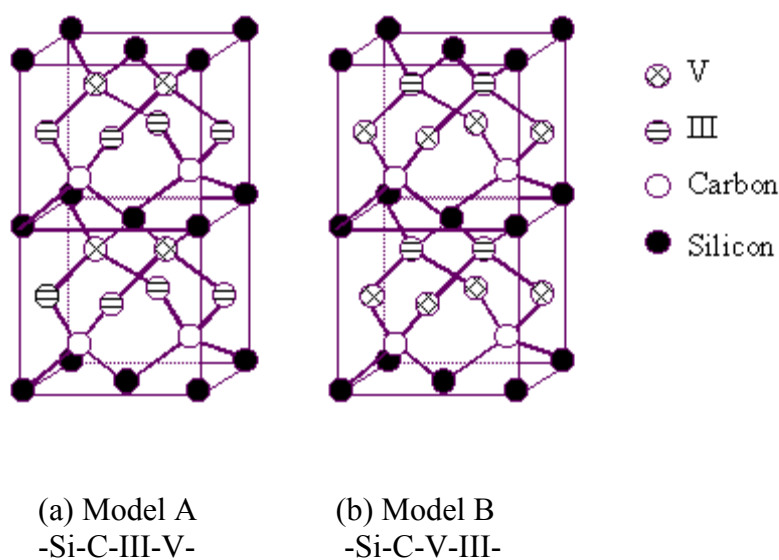

(a) Model A
-Si-C-III-V-

(b) Model B
-Si-C-V-III-

⊗ V
⊖ III
◯ Carbon
● Silicon

Fig 1. J. C. Zheng et al.

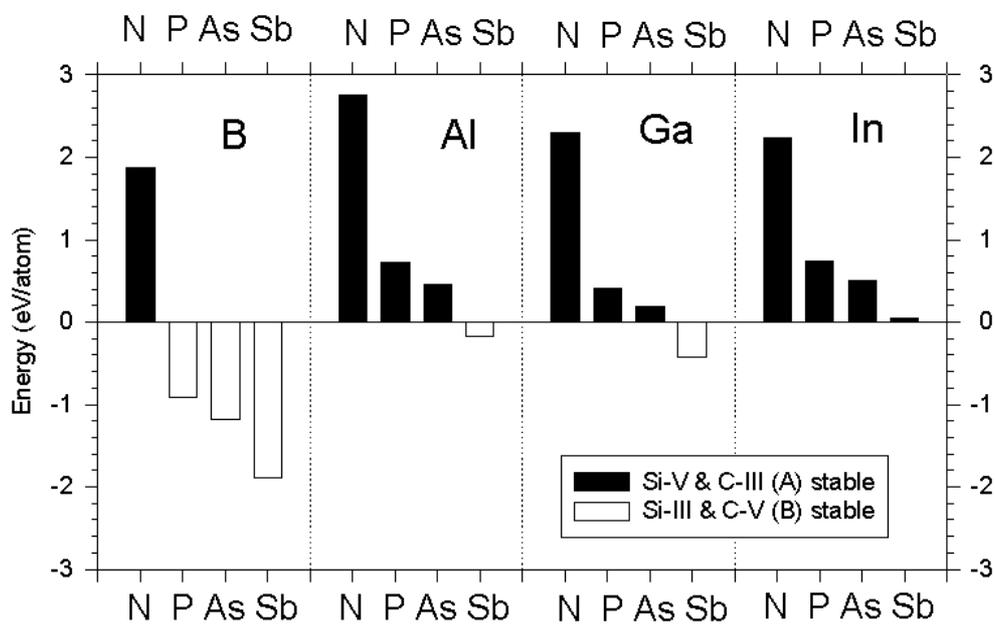

(a)

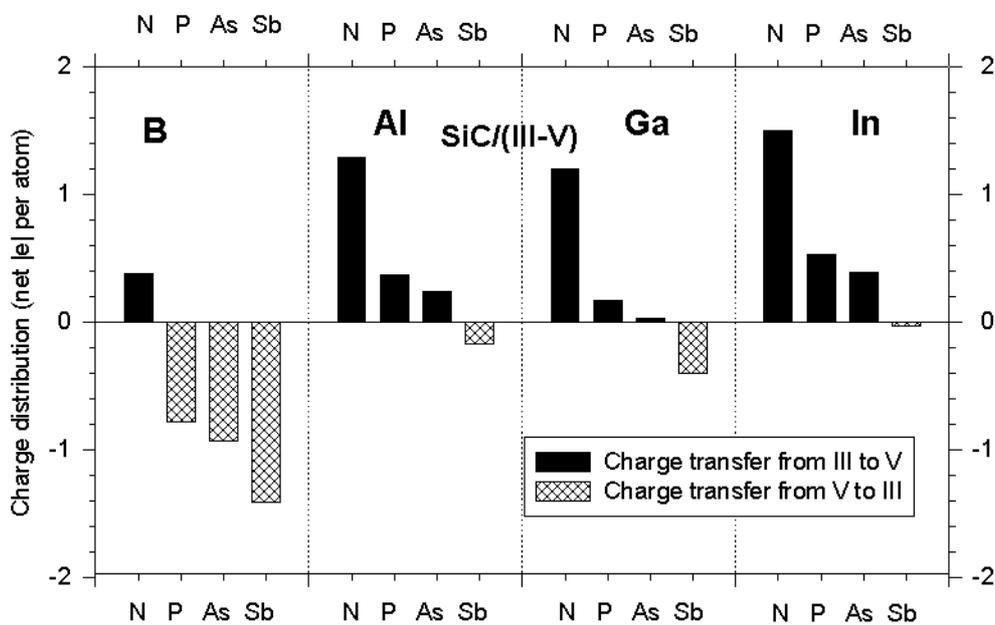

(b)

Fig. 2. J. C. Zheng  et al.

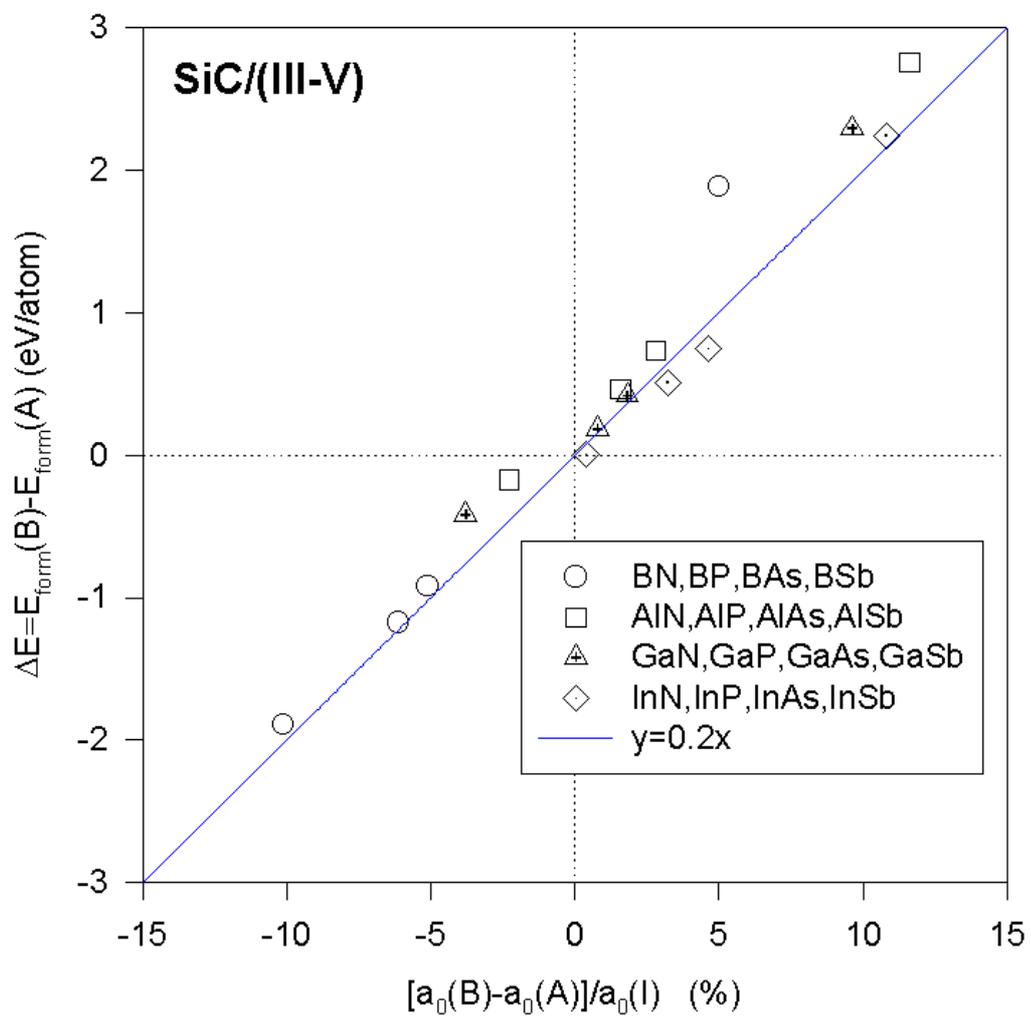

Fig. 3.  J. C. Zheng et al